\documentstyle[11pt]{article}
\def\th{\theta}

\def\d{\delta}

\def\m{\mu}

\def\o{\omega} \def\O{\Omega}

\def\be{\begin{equation}}
\def\ee{\end{equation}}
\def\bea{\begin{eqnarray}}
\def\eea{\end{eqnarray}}
\def\vA{\vec{A}}
\def\vH{\vec{H}}
\def\vE{\vec{E}}
\def\vL{\vec{L}}
\def\vr{\vec{r}}
\def\vY{\vec{Y}}
\def\vj{\vec{\jmath}}
\def\vnab{\vec{\nabla}}
\def\llra{\longleftrightarrow}

\begin{document}

\begin{flushright} BRX TH-570 \end{flushright}

\vspace{.4in}

\begin{center}
{\large\bf Multipole Radiation in Lorentz Gauge}

Howard J.\ Schnitzer\footnote{Research supported in part by the
DOE under grant DE-FG02-92ER40706\\ \hspace*{.2in}schnitzr@brandeis.edu}\\
Martin Fisher School of Physics\\
Brandeis University\\
Waltham, MA 02454
\end{center}

\noindent{\small {\bf Abstract}:  The multipole expansion for
electromagnetic radiation, valid for all wave-lengths and all
distances from bounded sources, is presented in Lorentz gauge,
rather than the usual Coulomb gauge.  This gauge is likely to be
preferred in applications where one wishes to maintain manifest
Lorentz invariance.  The presentation also serves as a useful
exercise in the use of vector spherical harmonics.}

\vspace{.1in}

The multipole expansion is a standard issue for the description of
electromagnetic radiation.  In many applications one may employ
suitable simplifications such as the long-wave limit, or
restriction to just the asymptotic behavior of the fields.  It is
of interest in other contexts to describe the multipole expansion,
valid for all wave-lengths and all distances from bounded sources.
Typically this expansion is presented in Coulomb gauge.  However
for applications where one wishes to maintain manifest Lorentz
invariance, the Lorentz gauge is preferable.  Of course, the
resulting electric and magnetic fields do not depend on gauge
choice. But in some applications, the multipole expansion in terms
of Lorentz covariant potentials is particularly useful.

It is the purpose of these notes to describe the multipole
expansion of electromagnetic radiation in Lorentz gauge valid for
all wave-lengths and all distances from bounded sources, as we are
not aware that this is available elsewhere.  Other useful
discussions, with other aspects of the subject, are to be found in
refs. [1--5].  Various useful properties of spherical Bessel
functions are available in standard references on quantum
mechanics, for example.

The electric and magnetic fields may be described in terms of
vector and scalar potentials $\vA$ and $\phi$, \be \vH (\vr ,t) =
\vnab \times \vA (\vr ,t)
 \ee
 and
 \be
 \vE (\vr ,t) = -\vnab \phi (\vr ,t) -{\textstyle{\frac{1}{c}}}\:
 {\textstyle{\frac{\partial}{\partial t}}} \: \vA (\vr ,t)\; .
 \ee
In Lorentz gauge, one has the gauge condition
 \be
 \vnab \cdot \vA +{\textstyle{\frac{1}{c}}}\:
 {\textstyle{\frac{\partial\phi}{\partial t}}} = 0 \; ,
 \ee
 which implies the potentials satisfy the wave-equations
 $$
 - \nabla^2 \vA (\vr ,t) +{\textstyle{\frac{1}{c^2}}}\;
{\textstyle{\frac{\partial^2}{\partial t^2}}}\; \vA (\vr ,t) =
{\textstyle{\frac{4\pi}{c}}}\: \vj (\vr ,t)
 $$
and
 \be
 - \nabla^2 \phi (\vr ,t) +{\textstyle{\frac{1}{c^2}}}\;
{\textstyle{\frac{\partial^2}{\partial t^2}}}\; \phi (\vr ,t) =
4\pi\rho (\vr ,t)
 \ee
in terms of the bounded current and charge densities $\vj$ and
$\rho$, respectively.

It is convenient to define the fourier transform in time of a
function $f (\vr ,t)$ by
 \be
 f_\omega (\vr ) ={\textstyle{\frac{1}{2\pi}}} \int dt \:
 e^{i\omega t} \: f(\vr ,t) \; ,
 \ee
 so that wave-equations are transformed to the Helmholtz
 equations, with $k = \omega /c$,
 $$
 - (\nabla^2 + k^2 ) \vA_\omega (\vr )
 ={\textstyle{\frac{4\pi}{c}}}\: \vj_\omega (\vr )
 $$
 and
 \be
 - (\nabla^2 + k^2 ) \phi_\omega (\vr )
 = 4\pi\rho_\omega (\vr ) \; .
 \ee
We solve for the potentials in terms of the sources, and then
reconstruct the electric and magnetic fields.

 The approach
described here has some overlap with that of Rose [1].

An arbitrary vector field, $\vA (\vec{r})$ can be expanded in
spherical waves \be \vA (\vr ) = \sum^\infty_{J=0} \:
\sum^J_{M=-J} \vA (J,M,\vr )\; .
 \ee
The expansion coefficients can be given in terms of vector
spherical harmonics.
 \be
\vA (J,M,\vr ) = {\textstyle{\frac{1}{r}}} \sum^{J+1}_{\ell = J-1}
\left[f_\ell (J,M;r) \vY^M_{J\ell 1}\right]
 \ee
where $f_\ell(J,M;r)$ is a function of the radial coordinate $r$,
and $\vY^M_{J\ell 1 }$ are the vector spherical harmonics defined
by
 \be
\vY^M_{J\ell 1} (\th , \phi ) = \sum^\ell_{m=-\ell} \:
\sum^1_{q=-1} (\ell m 1q | \ell 1 J M ) Y^m_\ell (\th , \phi )
e^q_1 \; .
 \ee
In (9) $e^q_1$ is a spherical unit vector, $Y^m_\ell (\th , \phi
)$ is the usual spherical harmonic satisfying $Y^{m*}_\ell =
(-1)^m Y^m_\ell$, and $(\ell m 1 | \ell 1 JM)$ is a
Clebsch--Gordan coefficient. If $e_x, \: e_y, \: e_z,$ are the
three rectangular unit vectors, then
\begin{eqnarray}
 e^1_1 & = & - \textstyle\frac{1}{\sqrt{2}} \; (e_x + ie_y) \nonumber \\
e^{-1}_1 & = &\textstyle\frac{1}{\sqrt{2}} \; (e_x - ie_y)\nonumber  \\
e^0_0 & = & e_z \nonumber \\[.1in]
e^{q*}_1 & = & (-1)^q e_{-q}\; .
\end{eqnarray}
The vector spherical harmonics obey the orthogonality
 \be
\int d\O \left[ \vY^{M{^*}}_{J\ell 1} (\th , \phi ) \cdot
\vY^{M'}_{J\ell ' 1} (\th , \phi ) \right] = \d_{JJ'} \d_{\ell\ell
'} \d_{MM'}\; .
 \ee
Using this one finds
 \be
\textstyle \frac{1}{r} \: f_\ell (J,M;r ) = \displaystyle{\int}
d\O
 \left[ \vY^M_{J\ell 1} (\th , \phi )\right]^* \cdot \vA (\vr )
 \; .
 \ee

In Lorentz gauge, the vector potential can be written in terms of
the Green's function
 \be
\textstyle \vA_\o(\vr ) = \frac{1}{c} \: \displaystyle{\int} \:
d^3r '  \: \vj_\o (\vr{\:'}) G_k
 (|\vr - \vr{\:'}|)  \hspace{.4in}\; , \hspace{.2in}
 k=\frac{w}{c}
 \ee
where $G_k (|\vr - \vr{\:'}|) = \frac{e^{ik|\vr - \vr '|}}
 {|\vr - \vr{\:'}|}$ .  Hence, for (13) one has
 \be
 \textstyle\frac{1}{r} \: f_\ell (J,M;r) = \frac{1}{c} \: \displaystyle{\int} d^3r'\vj_\omega
 \:  (\vr{\:'})\cdot \int d\O \left[ \vY^M_{J\ell 1} (\th , \phi )  \right]^* G_k (|\vr -
 \vr{\:'}|)\; .
 \ee

Consider the expansion of (14) for bounded sources, for which $r'
\leq R$ and $r \geq R$,
 \bea
 \int d\O\vY^{M*}_{J\ell 1} G_k (|\vr -  \vr{\:'}|) & = & \int d\O
 \sum^\ell_{m=-\ell} \; \sum^1_{q=-1} (\ell m 1 q|\ell  1 JM)^*
 Y^{m*}_\ell (\th , \phi ) e^{q*}_1 G_k (|\vr -
 \vr{\:'}|)\nonumber \\[.15in]
 & = & \sum^\ell_{m=-\ell} \; \sum^1_{q=-1}(\ell m 1 q|\ell  1 JM)^*
 e_1^{q*} \: \int d\O Y^{m*}_\ell (\th , \phi )
 \sum^\infty_{\ell ' = 0}\:
 \sum^{\ell '}_{m'= - \ell '}ik \; h^{(1)}_{\ell '}  (kr) \times \nonumber \\[.15in]
 &&  j_{\ell}
 (kr ')Y^{m'}_{\ell '} (\th , \phi)Y^{m'*}_{\ell '} (\th ',\phi ') \nonumber \\[.15in]
 & = & \sum^\ell_{m=-\ell} \; \sum^1_{q=-1}(\ell m 1 q|\ell  1 JM)^*
 e_1^{q*}
 Y^m_\ell (\th ',\phi ' )^*[ ik h^{(1)}_\ell (kr) j_\ell
 (kr')] \nonumber \\[.15in]
 & = & \vY^{M*}_{J\ell 1} (\th ', \phi ') ik \: h^{(1)}_\ell (kr)
 j_\ell (kr ') \; .
 \eea

In summary
 \be
 \left\{ \begin{array}{rcl}
  \frac{1}{r} f_\ell (J,M; r) & = & ik h^{(1)}_\ell (kr)
  \frac{1}{c} \displaystyle{\int} d^3 r' j_\ell (kr')\vj_\omega \, (\vr{\:'}) \cdot
  \vY^{M*}_{j\ell 1} (\th ', \phi ')   \\[.15in]
  \vA_\omega (J,M,\vr )& = & \frac{1}{r} \displaystyle{\sum^{J+1}_{\ell = J-1}} f_\ell (J,M;r)
  \vY^M_{J\ell 1} (\th , \phi )\\[.15in]
  \vA _\omega (\vr ) & = & \displaystyle{\sum^\infty_{J=0} \: \sum^J_{M=-J}} \vA_\omega (J,M,\vr )\; .
  \end{array}
  \right.
 \ee
  Putting this together, we can write
 \be
 \vA_\omega (\vr ) = \sum^\infty_{J=0} \: \sum^J_{M=-J} \:
 \sum^{J+1}_{\ell = J-1} ik h^{(1)}_\ell (kr) a^M_{J\ell 1}
 \vY^M_{J\ell 1} (\th , \phi )
 \ee
 where the coefficient
\be
 \textstyle a^M_{J\ell 1} = \frac{1}{c} \displaystyle{\int} d^3r' j_\ell (kr ') \left[
 \vj_\omega
 \,  (\vr{\:'}) \cdot \vY^{M*}_{J\ell 1} (\th ' , \phi ' )\right]
 \; .
\ee

 Similarly, for the scalar potential
 \be
 \phi_\o (\vr )= \int d^3 r' \rho_\o (\vr{\:'}) G_k (|\vr - \vr{\:'}|)
 \ee
 or
 \be
 \phi_\omega (\vr ) = \sum^\infty_{\ell = 0} \: \sum^\ell_{m=-\ell} ik
 h^{(1)}_\ell (kr) b^m_\ell Y^m_\ell (\th , \phi )
 \ee
 where
 \be
 b^m_\ell = \int d^3r' \rho (\vr{\:'})j_\ell (kr') Y^{m*}_\ell (\th ' , \phi ')
 \; .
 \ee
 The four coefficients $b^m_\ell$ and $a^M_{J\ell 1}$ are not
 independent.  They are related by the conservation of charge,
 which is satisfied by computing the Lorentz
 condition, which will also ensure the gauge invariance of our
 results for the fields $\vec{E}$ and $\vec{H}$.  So one must
 satisfy $\vnab \cdot \vA_\omega - i k \phi_\omega = 0$ in this gauge.
 Before doing this, it
 is convenient to tabulate some useful formulae.

 For any $\Phi (r)$
 $$
 \left\{ \begin{array}{lcl}
 \vnab \cdot \left[ \Phi (r) \vY^M_{J,J} (\th , \phi )\right] & = &
 0  \\[.15in]
 \vnab \cdot \left[ \Phi (r) \vY^M_{J,J+1} (\th , \phi )\right] & =
 & -
 \sqrt{\frac{J+1}{2J+1}} \: \left[ \frac{d}{dr} + \frac{J+2}{r}
 \right] \Phi (r) Y^m_\ell (\th , \phi ) \\[.15in]
 \vnab \cdot \left[ \Phi (r) \vY^M_{J,J-1} (\th , \phi )\right] & = &
 \sqrt{\frac{J}{2J+1}} \: \left[ \frac{d}{dr} - \frac{J-1}{r}
 \right] \Phi (r) Y^m_\ell (\th , \phi )
 \end{array} \right.
 $$
 $$
 \left\{ \begin{array}{lcl}
 \vnab \times \left[ \Phi (r) \vY^M_{J,J+1} \right] & = & i \left[
 \frac{d}{dr} + \frac{J+2}{r} \right] \Phi (r)
 \sqrt{\frac{J}{2J+1}}\: \vY^M_{J,J}\\[.15in]
 \vnab \times \left[ \Phi (r) \vY^M_{J,J-1} \right] & = & i \left[
 \frac{d}{dr} - \frac{J-1}{r} \right] \Phi (r)
 \sqrt{\frac{J+1}{2J+1}}\: \vY^M_{J,J}\\[.15in]
\vnab \times \left[ \Phi (r) \vY^M_{J,J} \right] & = & i \left[
 \frac{d}{dr} - \frac{J}{r} \right] \Phi (r)
 \sqrt{\frac{J}{2J+1}}\: \vY^M_{J,J+1}\\[.15in]
 && + \; i \textstyle{\left[ \frac{d}{dr} +
\frac{J+1}{r}\right]} \Phi (r) \sqrt{\frac{J+1}{2J+1}} \:
\vY^M_{J,J-1}
 \end{array} \right.~~~~~
 $$
 \be
 \begin{array}{rcl}
~~\vnab \left[\Phi (r) Y^m_\ell (\th , \phi )\right] & = & - \;
\sqrt{\frac{\ell
 + 1}{2\ell + 1}} \left( \frac{d}{dr} - \frac{\ell}{r} \right)
 \Phi (r) \: \vY^m_{\ell , \ell +1} (\th , \phi ) \nonumber \\[.15in]
 & & + \; \sqrt{\frac{\ell}{2\ell + 1}} \left( \frac{d}{dr} + \frac{\ell +1}{r} \right)
 \Phi (r) \: \vY^m_{\ell , \ell -1} (\th , \phi )\; .
\end{array}
\ee

Now we can use the following properties for {\it any} spherical
Bessel function ${z}_\ell (\rho )$.
\be
 \left\{ \begin{array}{lcl}
 {z}_{\ell -1} (\rho ) + {z}_{\ell +1} (\rho ) & = &
 \frac{2\ell +1}{\rho} \: {z}_\ell (\rho ) \\[.15in]
 \frac{d}{d\rho} \: {z}_\ell (\rho ) & = & \frac{1}{2\ell + 1}
 \: \left[ \ell {z}_{\ell -1} (\rho ) - (\ell + 1) {z}_{\ell +1} (\rho
 )\right]\; .\end{array} \right.
 \ee
 From these properties, we find
$$
 \left\{ \begin{array}{lcl}
 \vnab \cdot \left[ {z}_{J+1} (kr) \vY^M_{J,J+1} \right] & = & -
 k \: \sqrt{\frac{J+1}{2J+1}} \: {z}_J (kr) Y^M_J (\th , \phi ) \\[.15in]
\vnab \cdot \left[ {z}_{J-1} (kr) \vY^M_{J,J-1} \right] & = & -
 k \: \sqrt{\frac{J}{2J+1}} \: {z}_J (kr) Y^M_J (\th , \phi )
\end{array} \right.~~~
$$
$$
 \left\{ \begin{array}{lcl}
 \vnab \times \left[ {z}_{J+1} (kr) \vY^M_{J,J+1} \right] & = &
 ik \: {z}_J (kr) \sqrt{\frac{J}{2J+1}} \: \vY^M_{J,J} \\[.15in]
\vnab \times \left[ {z}_{J-1} (kr) \vY^M_{J,J-1} \right] & = &
 -ik \: {z}_J (kr) \sqrt{\frac{J+1}{2J+1}} \: \vY^M_{J,J} \\[.15in]
\vnab \times \left[ {z}_J (kr) \vY^M_{J,J} \right] & = &
 ik \left[ \: -{z}_{J+1} (kr) \sqrt{\frac{J}{2J+1}} \: \vY^M_{J,J+1} \right.\\[.15in]
&& + \left. \: {z}_{J-1} (kr) \textstyle{\sqrt{\frac{J+1}{2J+1}}
\: \vY^M_{J ,J-1}} \right]
\end{array} \right.
$$
 \be
 \vnab \left[ {z}_\ell (kr) Y^m_\ell \right] ~~ = ~~ k \textstyle{
 \left[ \sqrt{\frac{\ell + 1}{2\ell +1}} \: {z}_{\ell +1} (kr)
 \vY^m_{\ell , \ell +1} + \sqrt{\frac{\ell}{2\ell +1}} \:
 {z}_{\ell -1} (kr)  \vY^m_{\ell , \ell -1} \right]} \; .
 \ee

 Using these equations, we find that the Lorentz condition
 implies, writing $a_{J\ell 1}$ as $a_{J,\ell}$,
 \be
 b^m_\ell = i\textstyle{ \left\{ a^m_{\ell , \ell +1} \sqrt{\frac{\ell
 +1}{2\ell +1}} + a^m_{\ell , \ell -1} \sqrt{\frac{\ell}{2\ell +1}}
  \, \right\} } \; .
 \ee

The magnetic field is $\vH_\omega = \vnab \times \vA_\omega$.
Direct computation shows that
 \bea
  \vH_\omega & = &  \sum^\infty_{J=0} \: \sum^J_{M=-J} ik^2 \left\{ i
 h^{(1)}_J (kr) \vY^M_{J,J} \left[ \textstyle{ \sqrt{\frac{J}{2J+1}}} \:
 a^M_{J,J+1} -\sqrt{\frac{J+1}{2J+1}} \:  a^M_{J,J-1} \right] \right. \nonumber \\[.15in]
 & + & \left. i \: a^M_{J,J} \left[- \textstyle{ \sqrt{\frac{J}{2J+1}} }\:
 h^{(1)}_{J+1} (kr) \vY^M_{J,J+1}
 + h^{(1)}_{J-1} (kr) \sqrt{\frac{J+1}{2J+1}} \: \vY^M_{J,J+1}
 \right]\right\}\; .
\eea
 The electric field is constructed from
  $\vE_\omega = - \vnab \phi_\omega + ik \vA_\omega$.

Again by direct computation, and using (25), we have
 \bea
  \vE_\omega & = & (ik)^2  \sum^\infty_{J=0} \: \sum^J_{M=-J}
  \left\{ \left[ \textstyle{ \sqrt{\frac{J}{2J+1}}} \:
 a^M_{J,J+1} -\sqrt{\frac{J+1}{2J+1}} \:  a^M_{J,J-1}\right] \right.  \nonumber
 \\[.15in]
 & \times &  \left[ h^{(1)}_{J+1} (kr)  \textstyle{ \sqrt{\frac{J}{2J+1}} }\:
  \vY^M_{J,J+1} - h^{(1)}_{J-1} (kr) \sqrt{\frac{J+1}{2J+1}} \: \vY^M_{J,J-1}
 \right] \nonumber \\[.15in]
& + & \left. a^M_{J,J} \: h^{(1)}_J (kr)\vY^M_{J,J} \right\} \; .
 \eea
 One can verify that all the Maxwell equations are satisfied.

 We can define the magnetic multipole $\m^M_J \equiv a^M_{J,J}$,
 and the electric multipole
 \be
 p^M_J \equiv \textstyle{ \left[ \sqrt{\frac{J}{2J+1}} \: a^M_{J,J+1} -
 \sqrt{\frac{J+1}{2J+1}} \: a^M_{J,J-1} \right] }\; .
 \ee
 With these multipole coefficients, the fields take on a more
 compact, and symmetric appearance.
\be
 \left\{ \begin{array}{lcl}
 \vH_\omega & = & (ik)^2 \displaystyle{\sum^\infty_{J=0} \sum^J_{M=-J}} \left\{ p^M_J
 h^{(1)}_J (kr) \vY^M_{J,J} \right. \\[.2in]
  &  & +\left.\m^M_J \left[ -\sqrt{\frac{J}{2J+1}} \: h^{(1)}_{J+1} (kr) \vY^M_{J,J+1}
 + h^{(1)}_{J-1} (kr) \sqrt{\frac{J+1}{2J+1}}
 \vY^M_{J,J-1}\right]\right\} \\[.15in]
  \vE_\omega & = & (ik)^2 \displaystyle{\sum^\infty_{J=0} \sum^J_{M=-J}} \left\{ p^M_J
 \left[ \textstyle{ \sqrt{\frac{J}{2J+1}}} h^{(1)}_{J+1} (kr)\vY^M_{J,J+1} -  h^{(1)}_{J-1} (kr)
\sqrt{\frac{J+1}{2J+1}}\vY^M_{J,J+1}\right] \right.\\[.2in]
 &  & + \left. \m^M_J h^{(1)}_J (kr)\vY^M_{J,J}
 \right\}\; .\end{array}\right.
 \ee
Note that if $p^M_J \llra \m^M_J$, then
$$
\left. \begin{array}{lclc}
 \vE \; {\rm (electric)} & \llra & - \vH \; {\rm (magnetic)} \\
 \vE \; {\rm (magnetic)} & \llra & + \vH \; {\rm (electric)}
 \end{array} \right]
 \begin{array}{ll}
 \mbox{where the coefficients}\; \mbox{of} \; p^M_J \: \mbox{and} \; \m^M_J \: \mbox{are denoted}  \\
 \mbox{(electric) and (magnetic) respectively}.
 \end{array}
$$
 One can put some of these terms in a more familiar form, if one
 notes that
 \be
 \begin{array}{rcl}
 p^M_J & = & \left[\textstyle{ \sqrt{\frac{J}{2J+1}}  \: a^M_{J,J+1} -
\sqrt{\frac{J+1}{2J+1}} \: a^M_{J,J-1}} \right] \\[.2in]
& = & \frac{1}{c}\displaystyle{\int} d^3r' \vj_\omega (\vr{\:'})
\cdot \left\{\textstyle{ \sqrt{\frac{J}{2J+1}} \: j_{J+1} (kr')
\vY^{M*}_{J,J+1} - \sqrt{\frac{J+1}{2J+1}} \: j_{J-1} (kr')
\vY^{M*}_{J,J-1}}\right\}\\[.2in]
& = & \frac{-i}{ck} \displaystyle{\int} d^3r' \vj_\omega
(\vr{\:'}) \cdot \vnab \times [j_J (kr') \vY^{M*}_{J,J} ] \\
[.2in]
 & = & \frac{-i}{ck} \displaystyle{\int} d^3r' [j_J (kr')
 \vY^{m*}_{J,J}] \cdot [\vnab \times \vj_\omega (\vr ) ] \; .
\end{array}
\ee
 One can prove the identity
 \be
\textstyle \hat{r} Y^m_\ell = \frac{\vr}{r}\: Y^m_\ell = -
\textstyle{\sqrt{\frac{\ell + 1}{2\ell +1}}}\:\vY^{m}_{\ell , \ell
+1} + \textstyle{ \sqrt{\frac{\ell}{2\ell +1}}}\:\vY^{m}_{\ell
,\ell -1}\; .
 \ee

Using this, and the previously noted identities
 \be
\textstyle \vnab \times \left[ j_J (kr) \vr Y^M_J\right] =
\frac{i}{r} \sqrt{J(J+1)}\: j_J (kr) \vY^M_{J,J}\; .
 \ee
Hence
 \be
\begin{array}{rcl}
\m^M_J & = & \textstyle{\frac{-i}{c}} \displaystyle \int d^3r'
\vj_\omega (\vr{\:'}) \cdot \textstyle{\frac{r'}{\sqrt{J(J+1)}}}
\: \vnab \times \left[ j_J (kr') \textstyle{\frac{\vr{\:'}}{r '}}
\: Y^{M*}_J (\th ', \phi ')
\right] \\[.15in]
& = & \textstyle{\frac{-i}{c}} \displaystyle \int d^3r' [\vnab
\times (r' \vj_\omega (\vr{\:'}))] \cdot \left[ j_J
(kr')\textstyle{\frac{\vr{\:'}}{r '}} \:
Y^{M*}_J(\th ', \phi ')\right] \\[.15in]
& = & \textstyle{\frac{-i}{c}} \displaystyle \int d^3r' [\vnab
\times \vj_\omega (r' )] \cdot \vr{\:'} \left[ j_J (kr')
Y^{M*}_J(\th ', \phi ')\right] \; .
\end{array}
 \ee
Note that only the transverse component of the current enters as
the source, although in the longwavelength limit the longitudinal
component of the current is the source for the electric multipole
moments. This point is discussed in more detail by French and
Shimamoto [3] and by Snowdon [4]. Let us seek some alternate forms
for our multipole expansion. Note that $\vnab \times [{z}_J (kr)
\hat{r} Y^M_J] = \frac{i}{r} \sqrt{J(J+1)} \: {z}_J (kr)
\vY^M_{J,J}$, so that \be
\begin{array}{rcl}
{z}_J (kr) \vY^M_{J,J} & = & \textstyle{
\frac{-ir}{\sqrt{J(J+1)}}} \vnab \times \left[ {z}_J
(kr) \hat{r} Y^M_J \right] \\[.15in]
& = &\textstyle{\frac{-\vL}{\sqrt{J(J+1))}}}  \left[ {z}_J (kr)
 Y^M_J \right]
 \end{array}
 \ee
 where $\vL = -i \vr \times \vnab$.  Also
\be
\begin{array}{rcl}
\lefteqn{\left[ {z}_{J+1} (kr) \textstyle{ \sqrt{\frac{J}{2J
+1}}}\:\vY^{M}_{J ,J+1}  -
 {z}_{J-1} (kr) \textstyle{\sqrt{\frac{J+1}{2J +1}}}\:\vY^{M}_{J
 ,J-1}\right]}\\[.15in]
 & &=  \frac{1}{-ik} \: \vnab \times \left[ {z}_J (kr)\vY^{M}_{J,J}
 \right]\\[.15in]
 && = \frac{i}{k} \: \vnab \times \left\{ \textstyle{\frac{-\vL}{\sqrt{J(J+1}}}
 \: \left[ {z}_J (kr) Y^M_J \right] \right\} = \frac{-1}{k}
 \frac{\vnab \times \vL}{\sqrt{J(J+1)}} \:\left[ {z}_J (kr) Y^M_J
 \right]\; .
 \end{array}
 \ee
Using this
 \be
 \textstyle p^M_J = \frac{-i}{ck} \displaystyle{\int} d^3r'
\textstyle{\frac{\vL}{\sqrt{J(J+1))}}} \: \left[ j_J (kr')
Y^{M*}_J (\th ' , \phi ' ) \right] \cdot \left[ \vnab \times
\vj_\omega (\vr{\:'}) \right]
 \ee that is
 \be
\left\{
\begin{array}{rcl}
 p^M_J & = & \textstyle{\frac{-i}{ck\sqrt{J(J+1)}}} \: \displaystyle{\int} d^3r'
 \left[ \vnab \times \vj_\omega (\vr{\:'})\right] \cdot \vL \left[ j_J
 (kr') Y^{M*}_J (\th ', \phi ') \right] \\[.2in]
 \m^M_J & = & - \; \frac{i}{c} \int d^3r' \left[ \vnab \times
 \vj_\omega
 (\vr{\:'}) \right] \cdot \vr{\:'} \left[ j_J (kr{\:'}) Y^{M*}_J
 (\th ' , \phi ') \right]
 \end{array} \right.
 \ee
and
 \be
\left\{
\begin{array}{rcl}
 \vH_\omega & = & (ik)^2 \displaystyle{\sum^\infty_{J=0} \sum^J_{M=-J}}
 \left\{ \textstyle{\frac{p^M_J \vL}{\sqrt{J(J+1)}}}
 \left[h^{(1)}_J (kr) Y^M_J (\th , \phi ) \right] -
 \textstyle{\frac{\m^M_J \, i(\vnab \times \vL)}{k\sqrt{J(J+1)}}}
 \left[h^{(1)}_J (kr) Y^M_J (\th , \phi ) \right]\right\} \\[.2in]
 \vE_\omega & = & (ik)^2 \displaystyle{\sum^\infty_{J=0} \sum^J_{M=-J}}
 \left\{ \textstyle{\frac{ p^M_J i(\vnab \times \vL)}{\sqrt{J(J+1)}}}
 \left[ h^{(1)}_J (kr) Y^M_J (\th , \phi ) \right] +
 \textstyle{\frac{\m^M_J \, \vL}{\sqrt{J(J+1)}}}
 \left[h^{(1)}_J (kr) Y^M_J (\th , \phi ) \right]\right\}\; .
 \end{array} \right.
 \ee
Our results are now expressed in the same form as
 French and Shimamoto [3], as expected from the gauge invariance of the
 electric and magnetic fields.

\noindent{\bf References}

\begin{enumerate}
\item
 M.E.\ Rose, ``Multipole Fields," J.\ Wiley, 1955; ``Theory
of Angular Momentum," J.\ Wiley, 1957; 1961 Brandeis Lectures in
Theoretical Physics, vol.\ 2, W.A.\ Benjamin, 1962.
 \item
 A.R.\ Edmonds, ``Angular Momentum in Quantum Mechanics,"
 Princeton University Press, 1960.
 \item
 J.B.\ French and Y. Shimamoto, Phys.\ Rev.\ {\bf 91}, 898 (1952).
 \item
 S.C.\ Snowden, J.\ Math.\ Phys.\ {\bf 2}, 719 (1961).
 \item
 J.D.\ Jackson, ``Classical Electrodynamics," 3rd edition, J.\
 Wiley, 1999.
 \end{enumerate}

\end{document}